\documentclass[twocolumn,amsmath,amssymb,prb]{revtex4}

\usepackage{graphicx}
\usepackage{dcolumn}
\usepackage{bm}
\usepackage{upgreek}
\usepackage{url}
\usepackage{wasysym}
\usepackage{textcomp}

\begin{document}

\title{SPHERES,
J\"ulich's High-Flux Neutron Backscattering Spectrometer
       at FRM~II}

\author{Joachim Wuttke}
 \email{j.wuttke@fz-juelich.de}
\author{Alfred Budwig}
\author{Matthias Drochner}
\author{Hans K\"ammerling}
\author{Franz-Joseph Kayser}
\author{Harald Kleines}
\author{Vladimir Ossovyi}
\author{Luis Carlos Pardo}
 \altaffiliation[permanent address: ]{Grup de Caracteritzaci\'o de Materials,
Departament de F\'isica i Enginyeria Nuclear,
Universitat Polit\`ecnica de Catalunya, Barcelona, Spain.}
\author{Michael Prager}\thanks{deceased in 2008.}
\author{Dieter Richter}
\author{Gerald J.\ Schneider}
\author{Harald Schneider}
\author{Simon Staringer}
\affiliation{%
Forschungszentrum J\"ulich GmbH, 52425 J\"ulich, Germany}%

\begin{abstract}
SPHERES (SPectrometer with High Energy RESolution) is 
a third-generation neutron backscattering spectrometer,
located at the 20\,MW German neutron source FRM~II
and operated by the J\"ulich Centre for Neutron Science.
It offers an energy resolution (fwhm) better than $0.65$~$\upmu$eV,
a dynamic range of $\pm31$~$\upmu$eV,
and a signal-to-noise ratio of up to 1750:1.\\ ~\\
This preprint is outdated.
Please use only the published version:
\textbf{Rev. Sci. Instrum. 83, 075109 (2012);}
\url{http://dx.doi.org/10.1063/1.4732806};
also available from \url{http://apps.jcns.fz-juelich.de/spheres}

\end{abstract}

\maketitle

\section{Introduction}

Neutron backscattering
is a versatile technique for 
measuring the dynamics of spins, atoms, and molecules on a GHz scale.
Typical applications include 
hyperfine interactions in magnetic materials,
molecular rotations,
diffusion,
and relaxation in complex systems.\cite{Bee88,Hem00}

Historically, three instrument generations can be distinguished.
Proposed by Maier-Leibnitz in 1966,\cite{Mai66}
neutron backscattering was first realized 
in a test setup at the 4~MW reactor FRM
(Garching, Germany).
Count rates were of the order of one per hour,
and the signal-to-noise ratio
was 2:1 at best
so that it took several weeks to measure a single spectrum.\cite{AlBH69}
Following this demonstration of principle,
first-generation backscattering spectrometers
were build at J\"ulich (``$\pi$-Spektrometer''\cite{Ale72})
and Grenoble (IN10\cite{AlGH74,CoPH92,RaFG97},
IN13\cite{HeAP83}); 
they were reviewed in Refs.~\onlinecite{Spr80,Bee88,AlSH92,Dia92}.

In the second-generation instrument IN16 of Institut Laue-Langevin (ILL)
at Grenoble, consequent use has been made of focussing optics;
a deflector chopper allows to operate a spherical monochromator in
exact backscattering geometry.\cite{FrMB97b,FrGo01,Fri02}
Another instrument of this type is planned to be built at
the Australian research reactor OPAL.\cite{rsp:emu}

In third-generation spectrometers,
the focussing optics is made even more efficient by
a phase-space transform (PST) chopper.
This device, invented by Schelten and Alefeld in 1984,\cite{ScAl84}
enhances the neutron spectrum available at the monochromator,
and allows for a particularly compact instrument design.
The first realisation has been
the High-Flux Backscattering Spectrometer (HFBS)
at the National Institute of Standards and Technology
(NIST, Gaithersburg, Md.).\cite{GeNe98,MeDG03}

In this article, we present 
SPHERES, the SPectrometer for High Energy RESolution
built and operated by JCNS (J\"ulich Centre for Neutron Science)
at FRM~II (Forschungs\-neutronen\-quelle Heinz Maier-Leib\-nitz
of Technische Universit\"at M\"unchen at Garching, Germany).
A third PST based backscattering instrument is currently
under construction at the ILL (project IN16B)
where it will replace both IN10 and IN16.\cite{GoFr02,FrBS06}

The development of SPHERES has been documented
in several publications.\cite{KiPG99,KiPK00a,KiKP00b,KiKP01,KiKP02a,RoKP04}
First productive experiments were performed in June 2007,
and routine operation started in April 2008.
Since then,
two thirds of the beam time is offered to external users.
Experiment proposals are invited twice a year
through the JCNS user office.
By now (March 2012), nearly one hundred experiments have been performed,
and more than twenty publications have appeared.
In the following,
we describe the design of SPHERES,
its hardware and software components,
and its performance characteristics.
The supplemental material
contains a list of suppliers and some technical details.\cite{rsp:supp}

\section{Instrument Design}

\subsection{Instrument Layout}

In a backscattering spectrometer,
strong energy filtering of incident and scattered neutrons
is achieved by Bragg reflection from perfect monochromator and analyzer
crystals under angles close to 180$^\circ$.
This peculiar geometry leads to some complications in
the spectrometer design;
it entails the need for a primary beam deflector
and a duty-cycle chopper.
A particularly compact design is achieved
if these two functions are realized by one and the same
component: by a chopper that bears deflector crystals on its circumference.
As an important side-effect,
the Bragg deflection by moving crystals results in a PST
that enhances the useable flux at the monochromator.

\begin{figure}
  \centering\includegraphics[width=.46\textwidth]{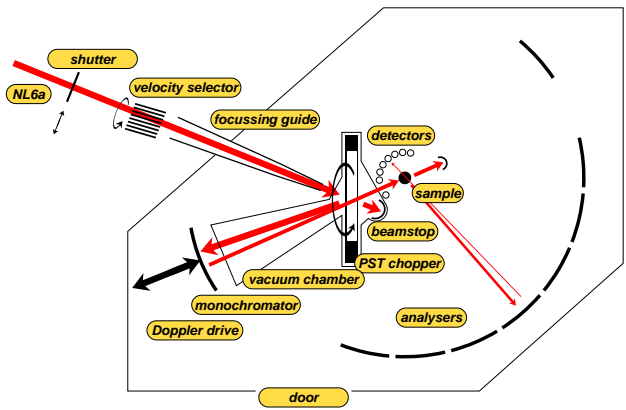}
  \caption{Basic layout of SPHERES.}
  \label{FLayout}
\end{figure}

The resulting instrument layout is shown in Fig.~\ref{FLayout}.
It is similar to the layout of HFBS,\cite{MeDG03}
with three major differences:
(1) HFBS is operated under vacuum whereas the SPHERES housing is gas tight
but not vacuum proof;
(2) at HFBS the monochromator is located outside the instrument housing;
(3) in the secondary spectrometer
detectors and analyzers have changed sides with respect to the primary beam,
resulting in a larger scattering angle range at SPHERES.

\subsection{Crystal Choices}

The basic design decision for a backscattering spectrometer
is the choice of the monochromator and analyzer crystals.
Initially, SPHERES was foreseen to be convertible
between a Si(111) and a Si(311) configuration.
Si(111) is the standard at most backscattering spectrometers.
The shorter wavelength of Si(311) gives access to a larger range
in the scattering wave number~$q$,
with promising application niches
in the study of diffusion on very short length scales.\cite{MeWP96a,SkCS98}
However, the lower resolution (which goes with $\lambda^{-2}$)
and background problems from Bragg reflections
(for instance from sample holders or from cryostat walls)
have so far prevented wide-spread useage.
In commissioning SPHERES,
the original mechanics had to be modified in many details
in order to block neutronic background channels and to provide
biological shielding.
As a result, the instrument has lost its convertibility;
the Si(311) option has been abandoned for the foreseeable future.

\begin{table}
  \begin{center}
  \begin{tabular}
    {@{}lcc@{}}
\hline\hline
  Bragg reflection for energy selection&Si(111) & Si(311)\\ 
  final neutron wavelength $\lambda_{\rm f}$ (\AA) & 6.271 & 3.275\\
  final neutron energy $E_{\rm f}$ (meV) &  2.080 &	 7.626 \\
  final neutron velocity $v_{\rm f}$ (m/s) &  630.9 & 1208 \\
  Bragg reflection for deflection&PG(002) & PG(004) \\
  deflection angle $2\alpha_0$ ($^\circ$) & 41.65 & 25.14 \\
  Schelten-Alefeld velocity $v_0$ (m/s) & 300.1 & 290.5 \\
  \hline\hline
  \end{tabular}
  \end{center}
\caption{\label{TSi}Parameters that depend on the choice of
Bragg reflections for energy selection and for deflection.
The current configuration of SPHERES uses silicon (111)
in monochromator and analyzers
and pyrolitic graphite (002) as deflector.}
\end{table}

Another important choice is pyrolitic graphite (PG) for
the deflector crystals.
Once deflector, monochromator, and analzyer crystals are chosen,
the neutronic parameters listed in Table~\ref{TSi} are fixed.
The angle $2\alpha_0$ between incoming and deflected beam
is obtained from the Bragg equation $2\cos\alpha_0=\lambda/(c/l)$
where
$c=6.710$~\AA\ is the lattice constant of~PG,
and $l$ the order of the Bragg reflection [00$l$].

\subsection{Time-of-Flight Logic\label{Stof}}

\begin{figure}
  \centering\includegraphics[width=.48\textwidth]{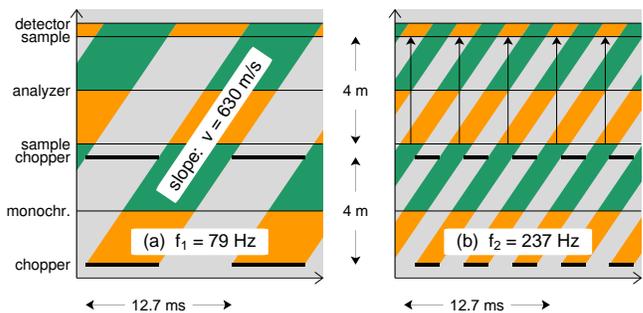}
  \caption{Neutron path versus time-of-flight: 
(a) As realized in SPHERES,
with chopper opening frequency $f_1=v_{\rm f}/4R$.
(b) Originally planned for SPHERES (and realized in HFBS),
with $f_2=3f_1$.
The upward arrows describe neutrons that are directly scattered
from the sample into the detector.}
  \label{Ftof}
\end{figure}

Another decisive hardware choice is the radius $R=2000$~mm
of the analyzer spheres.
This parameter fixes the time of flight
$\tau:=2R/v_{\rm f}=6.34$~ms for a round trip
from the sample to the analyzer and back.
During this round trip, the detectors see $n$ pulses of
directly scattered neutrons, interlaced with $n-1$ pulses
of energy-analyzed neutrons. 
If both types of pulses have the same duration $\tau_n$,
then $\tau_n=\tau/(2n-1)$, with $n=1,2,\ldots$
Accordingly, the chopper must open with a
frequency $f_n=1/(2\tau_n)$.

In the primary spectrometer,
neutrons are deflected by the closed chopper.
After a round trip to the monochromator and back,
they must hit the open chopper phase.
Therefore, the distance chopper -- monochromator $L_{\rm CM}$ must obye
$2L_{\rm CM}/v_{\rm f}=(2m-1)\tau_n$.
Since the monochromator should have about the same curvature as the analyzers,
the only practicable choice is $L_{\rm CM}=R$ and $m=n$.
All this is illustrated by the time-distance diagrams in Fig.~\ref{Ftof}.

HFBS uses $n=2$, and so will IN16B.
The present chopper of SPHERES also was originally designed for $n=2$,
but because of mechanical instabilities it is only operated with $n=1$.
This reduces the efficiency of the PST, 
but it has a positive effect upon the signal-to-noise ratio,
to be explained below (Sect.~\ref{Stofexp}).

\section{Instrument Components}

\subsection{Neutron Guide and Velocity Selector}\label{Sguide+nvs}

SPHERES is located in the neutron guide hall West of FRM~II.
Like all instruments in this hall,
it receives neutrons through beam tube~SR1
from the cold source.
SPHERES uses an end position at the secondary neutron guide NL~6a.
This guide has a cross-section of $60\times120$~mm$^2$;
its curved segments have a radius of 1000~m.
The guide is shared with two upstream instruments, MIRA\cite{rsp:mira} and DNS,
that use thin crystal monochromators to extract neutrons at shorter wavelengths
than those needed at SPHERES.

The instrument shutter of SPHERES and its mechanical neutron velocity selector
are located in two separate biological shieldings.
The velocity selector (EADS Astrium) reduces the incoming neutron spectrum
to about 6.27$\pm0.34$~\AA.
It is followed by a convergent neutron guide 
(``anti-trumpet'', by S--DH)
that reduces the beam section by
a factor of~11.5 to $25\times25$~mm$^2$.
Neutrons leave this guide system with a divergence
of about $\pm47$~mrad horizontal, $\pm64$~mrad vertical.
For details on all this, 
see the Supplemental Material.\cite{rsp:supp}

\subsection{Phase-Space Transform Chopper}

\begin{figure}
  \centering\includegraphics[width=.37\textwidth]{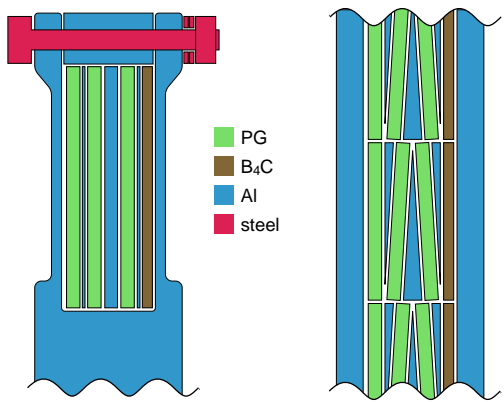}
  \caption{Deflector crystal assembly in the PST chopper.
In both drawings, the neutrons come from the left.
Left drawing: cross section.
Right drawing: view from outside towards the chopper axis
(the closing Al element and the screws are not shown).
The deflector stacks consist of three pyrolitic graphite mosaic crystals,
tilted by aluminum wedges, and followed by a $^{10}$B$_4$C absorber plate.}
  \label{FchopR}
\end{figure}

The phase-space transform (PST) chopper is the key component
of a third-generation backscattering spectrometer.
As said above, it combines the functionalities of a beam deflector
and of a duty-cycle chopper.
Half of its circumference 
bears deflector crystals that redirect incoming neutrons towards the
monochromator.
The other half circumference is open.
When the chopper is open,
neutrons coming out of the neutron guide
are transmitted towards a beam stop that is integrated in the chopper housing,
and neutrons coming back from the monochromator are transmitted
towards the sample.

To enable spectral scans,
a sufficiently large wavelength band
must be forwarded towards the monochromator.
This requires deflector crystals with large mosaicity.
Pyrolitic graphite (PG) crystals,
specified for a mosaicity of 2.5$^\circ$ fwhm
(Advanced Ceramics)
were selected according to rocking curves
measured one-by-one on the triple-axis neutron spectrometer
UNIDAS at J\"ulich.
To further enhance the horizontal mosaicity,
the crystals are assembled in stacks of three, with tilts of
$0^\circ$, $+2.5^\circ$, and $-2.5^\circ$ imposed by aluminum wedges
(Fig.~\ref{FchopR}).

Intensity and width of the forwarded neutron band are further enhanced
by the horizontal motion of the deflector crystals.
This phase-space transform `from white to wide'
has first been suggested 
in a conference proceedings of limited circulation;\cite{ScAl84}
an easily accessible summary has been provided in Ref.~\onlinecite{MeDG03}.
The Bragg reflection by moving mosaic crystals
compresses the energy distribution and thereby
enhances the flux in the acceptance range
of the backscattering monochromator.
In accordance with Liouville's theorem,
the compression of the energy distribution
is accompanied by a widening of the angular distribution.
This is not a severe drawback
as long as the angular resolution of the entire instrument
is limited by that of the secondary spectrometer.

Assuming an almost collimated incoming beam,
an optimum transform is predicted
for a deflector speed of
\begin{equation}
  v_0 = v_{\rm f}\frac{\sin\alpha_0}{\cos2\alpha_0}
\end{equation}
with the neutron velocity~$v_{\rm f}$ and the PG deflection angle~$\alpha_0$
as given in Tab.~\ref{TSi}.
For Si(111) and PG(002), one obtains $v_0=300.1$~m/s.
In the meantime, several simulations\cite{MeDG03,HeFS11,rsp:simW}
and a measurement\cite{MeDG03}
indicate that under realistic conditions the optimum deflector speed
is only about 250~m/s.

\begin{figure}
  \centering\includegraphics[width=.46\textwidth]{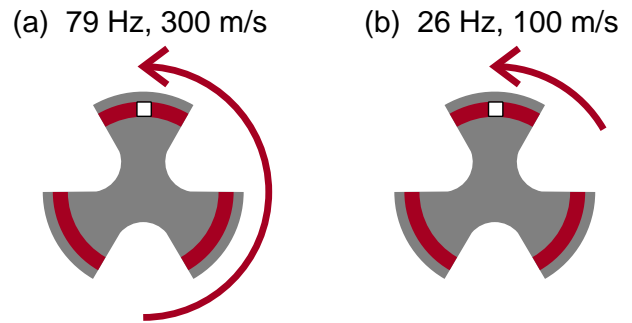}
  \caption{Chopper operation mode,
(a) originally intended, (b) currently realized.
The white square shows the illuminated area.
The arrows indicate the rotation during $\tau=6.34$~ms;
in this time, deflected neutrons travel from the chopper
to the monochromator and back.}
  \label{Ftof4b}
\end{figure}

The current chopper rotor has a radius of 650~mm;
the crystals are located at a mean radius of~$R=600$~mm.
The rotor has $k=3$ crystal bearing wings of 60$^\circ$ each.
Originally, the chopper was intended to operate with
a pulse fragmentation index of $n=2$ (Fig.~\ref{Ftof4b}a):
Neutrons that have made the round trip chopper -- monochromator -- chopper
would hit the second but next opening following the wing by which
they had been deflected.
The rotation frequency $f_{\rm rot}=(2n-1)f_n/k=f_2=78.9$~Hz
would result in a deflector speed of $2\pi R f_{\rm rot}=297$m/s.

Tensile forces for such $R$ and $f_{\rm rot}$ are considerable.
To keep them within admissible limits,
the high-strength aluminum alloy \textit{alumold}
(ALZNMg2Mn) was chosen,
and the geometry of the disk was optimized with finite-element calculations,
resulting in a construction with three spokes and two recesses
for each of the three wings.\cite{PrKK05}

Thermal instabilities of the bearings
and mechanical resonances
prevent us from running the 
current rotor with the intended 78.9~Hz.
As a fallback solution,
we are operating the chopper with $n=1$ and $f_{\rm rot}=f_1/3=26.3$~Hz
(Fig.~\ref{Ftof4b}b).
A new rotor is currently under construction;
it will have an asymmetric rotor ($k=1$) to achieve a
higher crystal speed
while maintaining~$n=1$.
As an engineering compromise,
a deflector velocity of 225~m/s has been chosen,
somewhat below the optimum, which, according to the simulations,
is quite flat.

\subsection{Monochromator and Analyzers}

In the standard configuration, 
the Si(111) reflection
is used in monochromator and analyzers.
Single-crystalline hexagonal wafers 
with a side length of 58.5~mm and a thickness of 500~$\upmu$m
were glued onto spherical supports
with a curvature radius of $R=2000$~mm.
The bending stress induces a lattice-constant gradient
that significantly improves the backscattered intensity
for a moderate cost in resolution.

The latter can be estimated quantitatively as follows.
The relative change of the lattice constant
at a distance $d$ from the neutral surface is $\kappa d/R$ 
with a material constant $\kappa=0.776$ for Si(111).\cite{StPo89}
Assuming a rectangular distribution of lattice constants,
and neglecting Darwin tails and other dynamical scattering effects,
one would expect a rectangular distribution
of backscattered energies with a full width of $\Delta E=0.81$~$\upmu$eV.
Convoluting this distribution with itself,
to account for monochromator and analyzer,
one obtains a triangular distribution with the same full width~$\Delta E$.
This is in reasonable agreement with the observed overall resolution
of SPHERES,
which is in first approximation a Gaussian with
a full width at half maximum of about 0.62~$\upmu$eV (Sect.~\ref{Sres}).

An alternate monochromator/analyzer set uses polished wafers
with a cut-in grid supposed to make them stress free.
The resolution is about 0.4~$\upmu$eV,
but the flux is so much lower that this configuration is of little use,
except possibly for the study of narrow tunneling or hyperfine splitting.

Additional monochromators offer the possibility of measuring with
a fixed energy offset.
A CaF$_2$(111) monochromator,
providing an
offset of 24~$\upmu$eV with respect to the Si(111) analyzers,
is already available.

The monochromator support,
made of carbon-fiber reinforced plastics,
has a size of 500$\times$250~mm$^2$
and a spherical curvature  of about $R+L_{\rm CS}/2=2180$~mm,
where $L_{\rm CS}=360$~mm is the distance between
the chopper and the nominal sample location,
so that the monochromator acts as a hollow mirror
that maps the chopper's PG crystals onto the sample.

The small-angle analyzers consist of eight massive aluminum rings
mounted on a common support.
Each ring has a width of 95~mm; between two rings there is a gap of 10~mm.
The innermost ring, with an inner radius of 165~mm,
proved useless because it receives fringe intensity from the direct beam;
therefore it was covered with Cd.
The other seven rings cover scattering angles~$2\vartheta$
from $9.0\pm1.4^\circ$ to $27\pm1.4^\circ$.
Typically, two rings focus into one detector.

The large-angle analyzers consist of seven shells,
made of cast aluminum
and finished by turning
They have a height of 2000~mm,
and a maximum width (in the scattering plane) of about 530~mm.
One shell covers a solid angle of about
$60^\circ\times15^\circ\simeq0.27$~rad$^2$.
The seven shells are located at
scattering angles from $30^\circ$ to $135^\circ$,
with only small gaps between them.
There is no one-to-one correspondence between individual shells and detectors;
the seven shells must rather be seen as one contiguous surface.

\subsection{Monochromator Doppler Drive}

A novel, resonance-free monochromator drive has been developed
jointly by Forschungszentrum J\"ulich and
an industrial company (Aerolas).
The development costs were shared with the ILL,
which received an identical drive for IN16 and for future use with IN16B.
The monochromator shell is mounted on an arbor that
is supported by two air bearing collars.
Driven by a linear motor, 
it performs a sinusoidal motion.
The displacement amplitude $Z$ can be chosen between $\pm25$ and $\pm75$~mm.
The maximum frequency is about $F=10$~Hz,
resulting in a maximum velocity amplitude of $V=2\pi FZ=4.7$~m/s.

To give an idea of the engineering difficulties,
let us note that the peak acceleration is $2\pi FV=295$~m/s$^2$.
The moving mass is nearly 10~kg,
and the maximum mechanical power is nearly 7~kW.
The linear motors generate a heat flux of up to 2.5~kW
that is removed by water cooling.
The peak current of each of the two motors is about 70~A at 600~V;
much of this electric power is recovered during deceleration.

We define the monochromator velocity~$v$
to take positive values if the monochromator is moving towards the sample.
In the rest frame of the monochromator,
backscattered neutrons have the velocity $v_{\rm f}=630.9$~m/s.
In the laboratory frame, they have the energy
\begin{equation}
  E_{\rm i}(v)=\frac{m_{\rm n}}{2}\left(v_{\rm f}+v\right)^2.
\end{equation}
The analyzers select scattered neutrons with energy $E_{\rm f}=E_{\rm i}(0)$.
In consequence, the \textit{sample energy gain} in the scattering process
is
\begin{equation}\label{Eomega}
   \hbar\omega=E_{\rm i}-E_{\rm f}
        =\left(2\frac{v}{v_{\rm f}}+\frac{v^2}{{v_{\rm f}}^2}\right)E_{\rm f}.
\end{equation}
The quadratic term causes a small asymmetry in the dynamic range.
With $V=4.7$~m/s, accessible energy transfers $\hbar\omega$
extend from $-30.9$ to 31.2~$\upmu$eV.

\subsection{Backscattering Detectors}

The detector bank for large scattering angles
consists of eleven counter tubes 24NH15,
bought from Eurisys,
filled with 4~bar of $^3$He.
They have a diameter of 1~inch
and a height of 210~mm.

The tubes are mounted vertically.
They are 
alternately arranged on two circles
with radii of 110 and 130~mm
at intervals $10^\circ$,
so that complete coverage is achieved
for an angular range of $110^\circ$ in the scattering plane.
In front of the tubes
(starting at a circle with radius 80~mm around the ideal sample centre)
there are radial, vertical Cd collimator plates.
The vertical opening of the detector block has been
reduced to about 80~mm in order to block neutrons that
do not come from the analyzers.

For small scattering angles,
front-window counters 10NH3/5X~P (also from Eurisys) are used,
filled with 3.5~bar of $^3$He.
Their diameter is 2~inch, the active length is 1~inch.
They are mounted around the exit nose of the chopper housing.

\subsection{Diffraction Detectors}\label{Sdiffhw}

SPHERES is equipped with position-sensitive detector tubes
(Reuter-Stokes)
for auxiliary diffraction measurements to be performed simultaneously
with the spectral scans.
The tubes are located in the horizontal scattering plane in
front of the analyzers;
inevitably, they shield a few percent of the effective analyzer surface.
At present, six tubes, with a diameter of 1~inch and
535~mm of active length,
provide coverage of scattering angles
from about 30 to $120^\circ$.

The resolution of this diffractometer is limited by the
horizontal spread of incident neutron directions,
which in the worst case is
$\pm\mbox{(half monochromator width)}/R=\pm250/2000=\pm12.5$\%.
This is acceptable because we do not aim for
flull-fledged structure determination
but only for monitoring structural changes.\cite{CoFL00}

\subsection{Instrument Housing with Argon Filling}\label{SAr}

The instrument housing has a surface of about 36~m$^2$,
and a volume of nearly 60~m$^3$.
Its walls and ceiling are made of steel plates of 5~mm thickness,
anchored to a steel frame, and proven to be gas tight.
Towards the inside, a 100~mm polyethylen layer 
moderates fast neutrons.
The innermost cladding of walls and ceiling consists of Cd plates
of 1~mm thickness.

The housing is reasonably gas tight and equipped
with pneumatics and controls so that it can be filled with argon.
This reduces losses by air scattering
in the secondary spectrometer,
where neutrons have to travel about 4.3~m through atmosphere.
In the primary spectrometer,
another 0.3~m are in atmosphere;
the remaining fligh path is in a vacuum chamber
that prolongates the chopper casing towards
the monochromator, as indicated in Fig.~\ref{FLayout}.
Displacing 90\% of the air by argon leads
to an intensity gain of more than 35\%,
in good accord with expectations from
tabulated neutron cross sections.
For technical details of the argon filling,
see the Supplemental Material.\cite{rsp:supp}

\subsection{Sample Environments}

The standard sample environment, used for most experiments,
is a cryostat SHI-950T (Janis)
It is a closed-cycle refrigerator system with cold head and compressor
manufactured by Sumitomo Heavy Industries.
In the low-temperature range from 4~K to 320~K,
the sample is cooled via exchange gas.
In the high-temperature range, from 290~K to 650~K
(700~K for limited amount of time),
the system acts as a cryofurnace, with the sample in vacuum.
The sample tube has a diameter of 60~mm;
samples are allowed to have diameters of up to 55~mm.

The standard sample stick is equipped with Si diodes DT670A
for temperatures from 4~K to 500~K.
A high-temperature stick with Pt~1000 thermometers
covers temperatures from 10~K to 700~K.
Gas-loading equipment is under preparation.

Sample environments are inserted and operated from
the platform on top of the instrument housing.
If the tail diameter exceeds 160~mm,
the secondary spectrometer must be temporarily realigned to a 
sample position slightly out of the focus of the monochromator.
To change the sample environment,
it is unavoidable to break the argon atmosphere.
Therefore,
experiments with non-standard sample environments are preferentially
scheduled at the beginning or at the end of a reactor cycle.
So far,
experiments have been performed with a dilution cryostat 
that gives access to temperatures down to 30~mK.
Other environments will be adapted and tried as requested by users.

\section{Instrument Control and Data Acquisition}

\subsection{High-Level User Interface}\label{Sgui}

During regular experiments,
physical access to the instrument is only needed for changing samples.
All other operations can be performed remotely.
A graphical user interface (GUI) allows
\begin{itemize}
\item to open and close the instrument shutter,
\item to start and stop the monochromator drive,
and to set its velocity amplitude,
\item to set the sample temperature, and to start temperature ramps,
\item to start and stop the data acquisition,
      to save files, and to reset histograms,
\item to submit, edit, start and stop experiment scripts.
\end{itemize}
It displays
\begin{itemize}
\item warnings and error messages,
\item the status of data acquisition, sample environment, and peripheral
      systems,
\item current count rates for all detectors,
      separately for direct, elastic, and inelastic scattering,
\item the latest acquired spectrum,
\item the latest acquired diffractogram,
\item the time evolution of various parameters in the last hours or days,
\item the status of experiment-script execution and the script commands ahead.
\end{itemize}

As the instrument is operated from different workplaces
(cabin in the neutron guide hall, office, home),
there is no simple way to maintain a traditional log book.
Therefore, we have replaced paper by digital storage,
using a simple wiki system \cite{s:dokuwiki}
as our \textit{instrument log}.
Besides, there are automatically generated 
\textit{event} and \textit{parameter} logs.

\subsection{Instrument-Control Daemons}\label{Sdaemon}

As a leading principle in the design of our instrument-control software,
we requested that no neutrons should be lost because of peripheral
software or network problems.
Therefore, the instrument's subsystems are controlled by
a number of independent background processes (daemons)
that continue to work even if the connection to upper-level processes is
interrupted.

The following daemons are the most important ones:
\begin{itemize}
\item spectral data acquisition (Sect.~\ref{Sdaq-sw}),
\item diffraction data acquisition (Supplemental Material)\cite{rsp:supp}
\item slow control daemon (Sect.~\ref{Sprofibus}),
\item temperature-controller controller,
      supervising the sample-environment temperature controller,
\item Doppler driver,
      supervising the control PC supplied by the manufacturer,
\item event logger, receiving log lines from other daemons 
      and writing them to an \textit{event log},
      thereby documenting
      user interventions, experiment-script execution steps,
      system warnings and error messages.
\item parameter logger,
      querying every 30~s some fifty status parameters from other daemons
      and writing them to a \textit{parameter log},
\item documentation server and raw-data postprocessor (Sect.~\ref{Sdoc}),
\item script engine, executing user-written experiment scripts,
\item GUI server, through which data and commands are
      channeled to and from the graphical user interface.
\end{itemize}
The spectral acquisition is time critical;
therefore it is written in the programming language~C,
and it is running on dedicated computer.
For all other processes,
computational speed is less important than clarity of code;
therefore they are written in an expressive scripting language, Ruby.

All computers are connected by ethernet (instrument LAN).
Interprocess communication (IPC) is based on
TCP (Transmission Control Protocol) messages.
Each daemon is listening on a TCP port
for commands given by other daemons.
For instance, the spectral acquisition daemon receives 
commands (start, stop, save, reset) from the GUI server,
and status queries from the parameter logger.

\subsection{Spectral Acquisition Electronics}\label{Sdaq-hw}

\begin{figure}
  \centering\includegraphics[width=.42\textwidth]{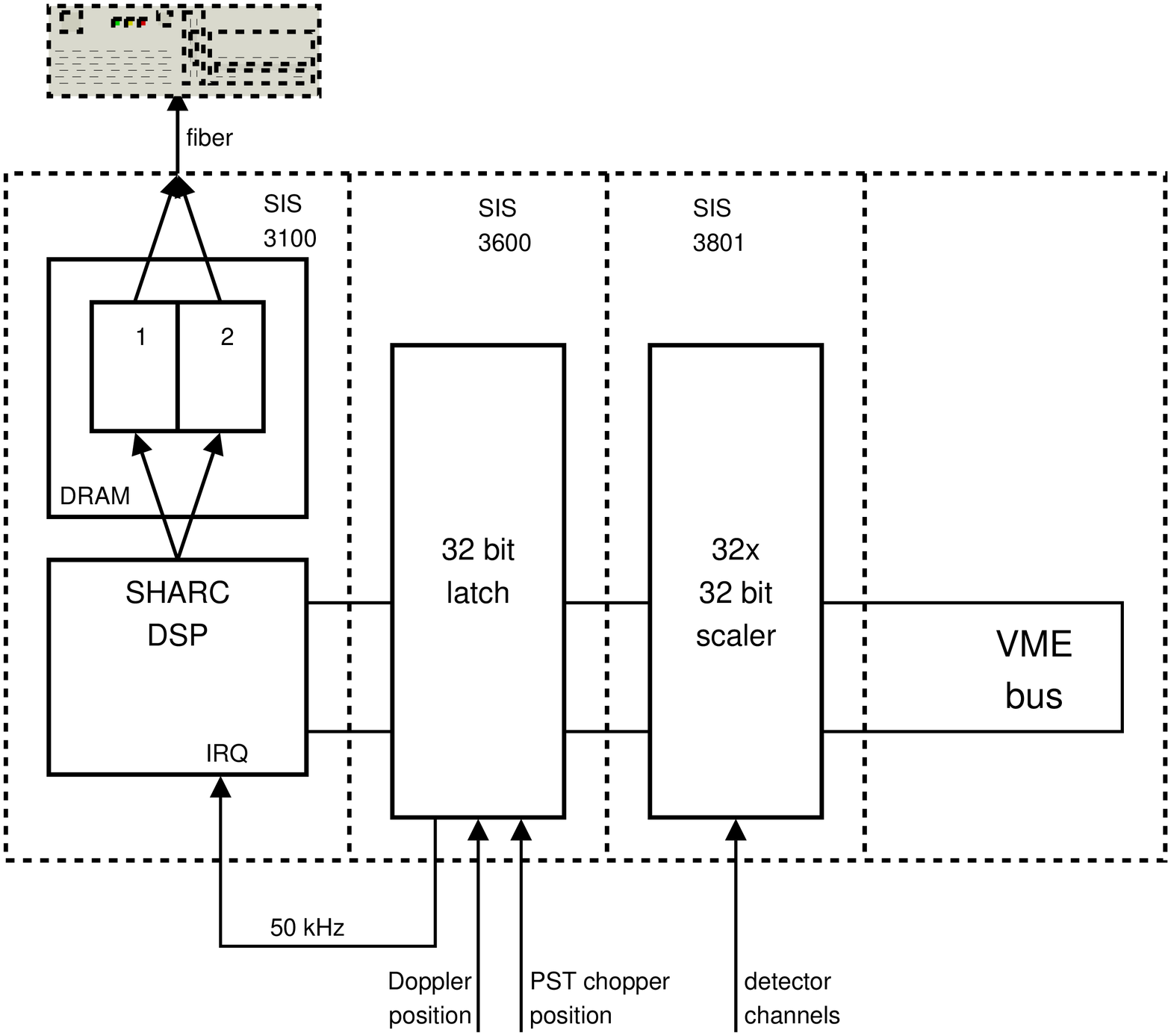}
  \caption{The data acquisition VME crate.\protect\cite{DrKK05}
The signal processor on the SIS~3100 sequencer board is a ADSP21061.
An optical fiber connects the crate to a SIS~1100 PCI card on the
dedicated data-acquisition PC.}
  \label{Fdaq}
\end{figure}

The data acquisition is done in a VME crate (Fig.~\ref{Fdaq})
using components and software from
Struck Innovative Systeme (SIS).\cite{DrKK05}
The input consists of TTL pulses
from the neutron-counter preamplifiers 
and from the optical encoders
sensing the translation of the monochromator Doppler drive (200 ticks per mm)
and the rotation of the PST chopper (20480 ticks per revolution).
The Doppler and chopper ticks are counted up and down in
home-made counter boards, then they are transmitted to a 32~bit latch.
The neutron counts are transmitted via a trivial home-made
interface board to a 32$\times$32 scaler.

The sampling period is $\tau=20$~\textmu s.
Once every~$\tau$,
a digital signal processor
reads the data from the latch and from the scaler,
writes them to DRAM, and resets the scaler.
The data form a fixed-length block of~20 32-bit words.
In the following, each block shall be identified by an index~$i$,
corresponding to a time $t_i=\mbox{const}+i\tau$.
One word encodes the Doppler position~$z_i$ and the chopper phase~$\chi_i$
(16 bits each);
the remaining 19 words~$n_{ji}$ accomodate the neutron counts
from up to 19~detectors~$j$.
The DRAM is divided into two buffers with 1'000'000 words each.
It takes 50'000 data blocks or 1~s to fill one buffer.
While one buffer is being filled,
the other buffer is read out via an optical link by the upstream PC.

\subsection{Spectral Acquisition Daemon}\label{Sdaq-sw}

The data-acquisition PC
(Opt\-er\-on 2$\times$2.4~GHz, 64bit),
is dedicated to running the spectral acquisition daemon.
It accumulates and stores histograms
containing neutron counts per energy channel and per detector.
These \textit{raw spectra}
are only a few correction steps (Sect.~\ref{Sslaw}) away from
the physical scattering law $S(q,\omega)$.

The daemon consists of an endless loop that waits for
one of the two DRAM buffer on the VME crate to be filled.
As soon as this is the case,
its contents is read via the optical link and pushed on a stack.
When the stack covers at least one Doppler oscillation period,
it is processed.
A sine function with four free parameters
(offset, amplitude, frequency, phase)
is fitted to the $z_i$,
using starting values retained from the previous fit.
Then, a loop over the data blocks~$i$ is performed,
to be described in the next three paragraphs.

Depending on the chopper phase $\chi_i$,
the data block is assigned a category $c(\chi_i)$,
which can be \textit{indirect}, \textit{direct},
or \textit{ambiguous},
depending on whether or not the neutrons have travelled the detour
sample--analyzer--sample or not.
In Sect.~\ref{Schopha} we will show 
how this categorization
is set up.
Ambiguous data blocks are not processed further.

To assign our scattering events to an energy channel~$\omega$,
we need to know the time
\begin{equation}\label{Eti}
  \hat{t}_i=
    t_i
    -(v_{\rm f}+v(\hat{t}_i))(L^0_{\rm i}-z(\hat{t}_i))
    -v_{\rm f} L^c_{\rm f}
\end{equation}
at which the neutron has been backscattered by the monochromator.
Here, $t_i$ is the detection time, and
$L_{\rm i,f}$ is the path travelled before and after scattering by the sample.
The path $L^c_{\rm f}$ depends on the category~$c$:
it is only about 10~cm in the direct case, but 4~m more in the indirect case.
The path offset $z(\hat{t}_i)$ and the velocity modulation $v(\hat{t}_i)$
are given by position and speed of the monochromator at time~$\hat{t}_i$.
This makes Eq.~(\ref{Eti}) nonlinear;
it is most easily solved by iteration.
For a given $\hat{t}_i$,
the analytic derivative of the fit function $z(t)$ is used to
obtain $v(\hat{t}_i)$.
Eq.~(\ref{Eomega}) then yields $\omega(v_i)$.
A quick estimate shows that the delay $\hat{t}_i-t_i$
is not negligible:
Neutrons in the \textit{indirect} category
travel more than 6~m from the monochromator to the detector,
which takes about 10~ms, corresponding to up to 1/10 of a Doppler period.

On starting the data acquisition for a given Doppler velocity amplitude,
an equidistant $\omega$ mesh has been chosen, with channels 
$[\omega_k-\Delta\omega/2,\omega_k+\Delta\omega/2]$.
This makes it easy to determine the pertinent channel
$k(i)=\lfloor (\omega(v_i)-(\omega_0-\Delta\omega/2))/\Delta\omega\rfloor$.
Finally, the neutron histograms 
of the pertinent chopper category $c(\chi_i)$ are incremented:
 $H^c_{jk}\leftarrow H^c_{jk}+n_{ji}$ for all detectors~$j$.
The time spent in energy channel~$k(i)$ is incremented accordingly:
$N^c_k\leftarrow N^c_k+1$.
Here ends the loop over~$i$.

All these computations must be completed
before the next DRAM buffer is overwritten.
After some optimizations,
this requirement could be fulfilled with sufficient margin:
In years of operation,
not a single data block has been lost because of computer overload.
As a side product,
a self-contained C library for Levenberg-Marquardt
least-squares minimization has been released
as an open-source project,\cite{s:lmfit}
serving practitioners from all areas of science and engineering.

Our histogramming relies on the assumption
that the motion of the monochromator
accurately follows a simple analytic function.
Analyses of recorded $z_i(t_i)$ traces have shown
that the motion accomplished by the Aerolas Doppler drive
is indeed almost perfectly sinusoidal,
deviations being negligible
on the scale set by the energy resolution of the spectrometer.

The accumulated histograms are written to files when
a preset time interval has elapsed or when a save command
is received through the daemon's IPC interface.
Files are saved to the local disk,
which is remote readable through NFS (Network File System).
The raw-spectra,
as other data produced by SPHERES,
are formatted as self-documenting YAML files.
YAML is a data-serialization specification
with intrinsic support for basic data structures like arrays and hashes
and with an emphasis on human readability;
parsers are available for all major programming languages.\cite{s:yaml}

\subsection{Safety Instrumentation and Slow Control}\label{Sprofibus}

SPHERES is equipped with a fail-safe programmable logic controller
(Fehlersichere speicherprogrammierbare Steuerung F-SPS,
Siemens Simatic S7 300F)
for safety instrumentation and slow control.\cite{KlSD05}
The F-SPS is connected via internal bus or Profibus-DP
to various sensors and actors.
A touchpanel next to the instrument door provides a status display;
it provides the only regular user interface for safety-critical
subsystems like the door lock and the argon pneumatics.

A simple daemon running on a compact-PCI computer couples the F-SPS
to the instrument LAN.
It has read access to all sensors and
write access to safety-uncritical actors,
including the shutter opener, which can be activated
if the instrument housing is closed and locked and a manual switch is
in ``remote control'' position.

\section{Data Processing}

\subsection{Raw-Data Postprocessor}\label{Sdoc}

The raw files produced by the spectral data-acquisition daemon
contain the energy mesh $\omega_k$,
the histograms $H^c_{jk}$ and $N^c_k$,
some information about the instrument and data-acquisition setup,
and the date and time when the measurement was started and stopped.
Had all other daemons crashed, this would be sufficient to
allow a valid data analysis.
In regular operation, however, users expect
data files to contain additional information like experiment title,
sample name, and sample temperature.
Since the spectral acquisition daemon is deliberately kept ignorant
about these things (not essential for its functioning,
and potential sources of bugs and failures),
the wanted information must be added ex post.

This is done by the raw-data postprocessor.
Running on another computer than the spectral acquisition daemon,
it reads a raw-spectrum file via NFS,
and looks up the start and stop time entries.
Then, it opens the parameter log,
reads the lines recorded during the spectral measurement,
and computes mean value and standard deviation of the sample temperature
and of some other parameters.
Textual information like experiment title and sample name,
usually entered by the user via the GUI,
is retrieved from another dedicated file system.
All this is written
to a user-friendly aggregated raw-data file.

\subsection{Raw-Data Reduction with SLAW}\label{Sslaw}

While some users want to work on the raw data,
most ask for standard software to deliver
normalized scattering laws $S(q,\omega)$.
To provide this service,
a new raw-data reduction program \textsc{Slaw} \cite{s:slaw} has been developed.
Compared to legacy software like SQW of the ILL,
\textsc{Slaw} offers better support for four-dimensional data sets
$S(q,\omega;T)$ or $S(q,\omega;t)$
obtained in time-resolved inelastic scans.
Also, \textsc{Slaw} does not bind users to a specific data format;
it supports a variety of output formats, and new output routines
can be easily added \textit{ad hoc}.

\textsc{Slaw} is controlled by an input script
that contains configuration commands
and a map that assigns raw scans to output files.
\textsc{Slaw} processes the raw data through the following steps
of which all but the first and the last are optional:
\begin{itemize}
\item 
    read raw data files, adding or subtracting subscans as needed;
\item
    perform binning in energy, scattering angle, or time slots,
    possibly eliminating some of these dimensions;
\item 
    subtract empty-can measurement,
    optionally weighted with a self-absorption factor;\cite{s:absco}
\item 
    normalize neutron counts $H^c_{jk}$
    to the measuring time $N^c_k\tau$ spent in 
    a given energy channel;
\item 
    correct for incident flux variations (Sect.~\ref{Smon});
\item 
    divide counts by elastic intensity of a reference measurement
    (usually vanadium or a low-temperature measurement with the current sample);
\item 
    save the scattering law.
\end{itemize}

\textsc{Slaw} has alternative input routines to process raw data
from other backscattering spectrometers.
Its structure is generic enough to 
allow future extension to time-of-flight spectrometers.

\subsection{Instead of a Monitor}\label{Smon}

Most inelastic neutron spectrometers
have a high-transmission neutron counter,
located a little upstream of the sample,
to monitor the incoming flux.
In SPHERES,
the windows of such a monitor would cause undesirable background.
We therefore prefer an indirect normalization.

We must account for two kinds of flux variation:
temporal variations of the reactor power, of the filling of the cold source,
and of the monochromator orientations of the upstream instruments,
and spectral variations of the primary spectrometer transmission.
The temporal variations of the incident beam
can be measured with good accuracy without inserting any device into the beam:
it is sufficient to monitor the $\gamma$ radiation emitted by 
the aluminum windows
at the end of the convergent guide and at the chopper entry.
Such a $\gamma$ monitor, still missing, will be installed soon.

The spectral variation of the flux at the sample is
determined by spectrum of the cold source, by the neutron guide system,
by the velocity selector, and most importantly,
by the PST chopper and the monochromator.
To make things more complicated,
the angle of the Bragg deflection in the PST chopper is energy dependent,
and the monochromator is moving in space so that the geometric-optical
focussing onto the sample is imperfect,
with a non-trivial interplay of energy, direction, and lateral spread.
To account for all this,
the spectral distribution of the incident flux is best measured
either by placing a neutron counter exactly at the sample position,
or by counting neutrons that are scattered from the sample directly
into the regular detectors.
The latter method is regularly used as the ``pseudomonitor'' of SPHERES.

\section{Performance}

\subsection{Flux and Count Rates}

The incident flux at the end of the convergent neutron guide
has been measured at two occasions with gold foil activation.
The results were
$(1.85\pm0.05)\cdot10^9$~$\text{cm}^{-2}\text{s}^{-1}$ in November 2005,
and $(1.19\pm0.05)\cdot10^9$~$\text{cm}^{-2}\text{s}^{-1}$ 
in February 2012.
A decrease of the order of 10\% was to be expected from
the two breaks in the upstream neutron guide (Sect.~\ref{Sguide+nvs}).
The remaining loss is not understood;
it must be feared that the neutron guides have degraded substantially.
To detect further degradation and to determine
necessary action, it will be important
to remeasure the flux more regularly in the coming years.

The flux at the sample position is more difficult to measure;
for details see the Supplemental Material.\cite{rsp:supp}
The best current estimate is $\phi_{\rm sam}=1.8\cdot10^6$~s$^{-1}$
within a a cross-section of $25\times40$~mm$^2$.
Here and in the following,
fluxes are reported for the open chopper state.
The average flux seen by a gold foil is only half of it.

For typical samples made for transmissions between 80\% and 90\%,
elastic count rates $\phi_{\rm el}$ are of the order 1000~s$^{-1}$.
Here as in the instrument's GUI display,
such numbers are per detector, and under the assumption
that all time is spent counting into this single channel.
The pseudomonitor rate $\phi_{\rm dir}$ of neutrons
that are directly scattered from the sample into the detector,
is about twice as large.
Reasons why $\phi_{\rm el}/\phi_{\rm dir}$ is smaller than~1 include:
solid angle mismatch (the detector block has a larger solid angle
than the analyzers, to ensure that all backscattered neutrons have
a chance to reach a detector),
absorption and scattering losses in the sample
(which must traversed a second time by backscattered neutrons),
losses along the 4~m from the detector to the analyzers and back,
imperfect energy selection in the monochromator,
finite backscattering probability in the analyzers.

\subsection{Dark Counts}

Dark count rates in each of the large-angle detectors are of the order of
0.02~s$^{-1}$ when the reactor is off,
and 0.1~s$^{-1}$ when the reactor is at 20~MW
and the intrument shutter closed.
This is negligible
compared to the background produced in our primary spectrometer.

When the instrument shutter is open and
the chopper is running,
but the exit window of the chopper (just before the sample position)
is closed by a Cd mask,
then the dark count rates in the large-angle detectors are
about 0.53~s$^{-1}$.
Only a small part of this background is due to slow neutrons
that penetrate from the primary into the secondary spectrometer,
circumventing the chopper exit window.
Most of the background is independent of the chopper phase,
which indicates an origin between the shutter and the chopper.
This has been confirmed by experimental tests.
Probably,
a n--$\gamma$ conversion in the supermirror coating of
the focussing neutron guide is followed by
a conversion to fast neutrons in the guide's lead shielding.

\subsection{Chopper Modulation of Count Rates}\label{Schopha}\label{Stofexp}

Along with the duty-cycle chopper,
a backscattering spectrometer needs a time-of-flight logic to
discriminate neutrons that are backscattered by the analyzers
from neutrons that are scattered directly from the sample into
the detectors.
In SPHERES, this time-of-flight discrimination is performed by
the spectral acquisition daemon (Sect.~\ref{Sdaq-sw}),
based on the chopper phase reading that is saved every 20~$\upmu$s
along with the Doppler position reading and the neutron counts.

\begin{figure}[htb]
  \centering\includegraphics[width=.42\textwidth]{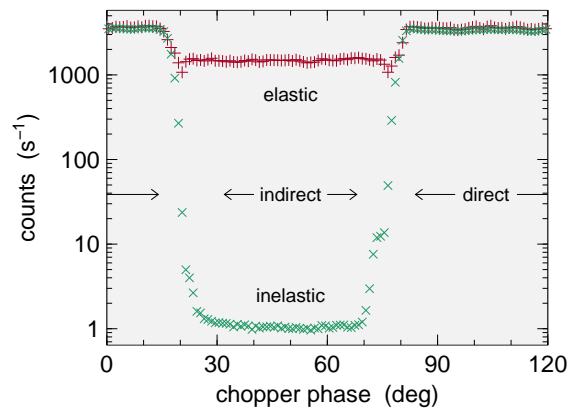}
  \caption{Neutron counts versus chopper phase.
Data from the same low-temperature
resolution measurement as in Fig.~\protect\ref{Fres}.
In the \textit{direct} phase,
neutrons are scattered without energy analysis from the sample
into the detectors.
In the \textit{indirect} phase,
registered neutrons come from the analyzer.
If the scattering is purely elastic, as in this example,
then the \textit{inelastic} signal is just undesired background.}
  \label{Fchopha}
\end{figure}

The parametrization of the discrimination must be determined empirically.
A chopper-phase histogram is saved in each raw data file.
This histogram is based on a rectangular grid
with 120 chopper-phase channels,
but only two energy channels,
to distinguish roughly between elastic and inelastic scattering.
Fig.~\ref{Fchopha} shows such data from a resolution measurement.
There is a range of nearly 60$^\circ$ with very high count rates,
independently of the Doppler velocity,
due to \textit{direct} scattering.
There is another range, labelled \textit{indirect},
where almost all neutrons have been backscattered by an analyzer,
as can be inferred from the strong difference
between the elastic and the inelastic channel.
And there are intermediate, \textit{ambiguous} ranges 
where directly scattered neutrons form an acceptable background
in the inelastic channels.
Part of this transitory range is due to the finite time
it takes the chopper to open and to close.
Other contributions include the following:
\begin{itemize}
\item In the present chopper,
the graphite crystals are kept in place by aluminum corner pieces.
When these pieces cross the primary neutron beam,
they cause incoherent scattering into $4\pi$,
thereby sending a diffuse neutron cloud into the secondary spectrometer.
\item Some directly scattered neutrons are retarded by phonon downscattering,
causing a sample-dependent extension of the chopper-closing phase.
\item Multiple scattering (first scattering in the sample, second scattering
by the sample environment or somewhere in the secondary spectrometer)
also presents a sample-dependent contribution
to the extended chopper-closing phase.
\end{itemize}

To parametrize the data acquisition daemon,
the instrument responsible must make a relatively arbitrary choice
which range of chopper phases shall be considered as \textit{indirect}.
In general,
users of SPHERES are more interested in an excellent 
signal-to-noise ration than in maximum count rates.
Therefore,
a relatively strict chopper-phase discrimination
that favors signal-to-noise ratio on the expense of total count rates
is preferred.
For most measurements in the past years,
we have set the \textit{indirect} interval to $35^\circ$,
which means that the duty cycle
(the fraction of time actually used for incrementing scattering histograms)
is no more than $35/120=0.29$.

For the future, one might consider a small modification
of the data acquisition code 
to simultaneously record energy histograms for
different chopper-phase discrimination settings.
Users could then decide \textit{ex post} whether 
to analyze
a ``strict'' data set with optimum signal-to-noise ratio,
or a ``loose'' data set with higher count rates.

\subsection{Resolution and Signal-to-Noise Ratio}\label{Sres}

The instrumental resolution can be empirically determined
by measuring the elastic scattering of a solid.
The scattering should be predominantly incoherent
to guarantee sufficient intensity at all scattering angles.
Vanadium, the elastic incoherent reference scattering \textit{par excellence},
is not well suited
because its absorption cross section at $\lambda=6.27$~\AA
is more than three times larger than its scattering cross section.
Therefore, 
we prefer a simple plastic foil 
(Hostaphan RN 250: a polyethylene terephthalate foil
of 0.25~mm thickness)
as a our reference scatterer.

In most experiments, however,
the resolution is measured on the user sample,
cooled to a low temperature where
almost all scattering is elastic.
This has the advantage that the sample geometry is exactly
the same as in the production measurements.

\begin{figure}[htb]
  \centering\includegraphics[width=.42\textwidth]{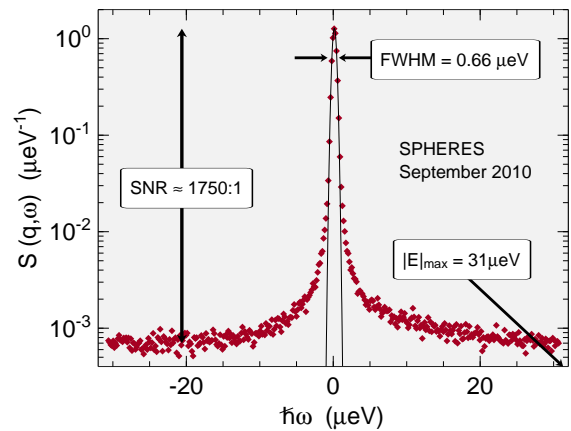}
  \caption{A resolution measurement with a user-provided sample
(a powder of hydrogen-rich organic crystals at 3~K,
in a flat cell calculated for 90\% transmission).
Cumulated data from the six large-angle detectors with the
best signal-to-noise ratio.
The solid line is a Gaussian fit to the central region
$|\hbar\omega|<0.6$~$\upmu$eV.}
  \label{Fres}
\end{figure}

Fig.~\ref{Fres} shows such a resolution spectrum from 
a user experiment.
The Doppler drive was running with the maximum velocity amplitude
of 4.7~m/s, resulting in a dynamic range of $\pm31$~$\upmu$eV.
To determine the resolution width and the elastic amplitude,
the centre of the elastic peak ($|\hbar\omega|<0.6$~$\upmu$eV)
has been fitted by a Gaussian,
which is an excellent approximation
though there are pronounced non-Gaussian wings 
at larger energy transfers.
The full width at half maximum (fwhm) is obtained by
multiplying the standard deviation of the Gaussian 
with $\sqrt{8\ln(2)}$.

When the Doppler drive is run with small to moderate velocity
amplitudes, the resolution at large scattering angles
is typically between 0.62 and 0.65~$\upmu$eV.
At the maximum Doppler frequency used here, the resolution is slightly reduced
(to 0.66~$\upmu$eV),
probably because of distortions of the monochromator support. 

The signal-to-noise ratio (SNR)
depends obviously on how we define the ``noise''.
If we average only over energies far off the central peak,
say $|\hbar\omega|\ge20$~$\upmu$eV,
then we obtain in the present example 1750:1.
The SNR also depends on the sample:
if the sample is too thin,
then the sample-independent constant background contributes relatively more;
if the sample is too thick,
then many backscattered neutrons are lost 
when they retraverse the sample.


\section{Comparative Status and Perspective}

In Table~\ref{TComp}, some design and performance parameters
of the three backscattering spectrometers IN16, HFBS and SPHERES are compared.
These parameters are intercorrelated as follows.

\begin{table}
  \begin{center}
  \begin{tabular}
    {@{}p{.43\columnwidth}p{.17\columnwidth}p{.17\columnwidth}p{.17\columnwidth}@{}}
\hline\hline
  instrument & IN16\footnote{With the \textit{unpolished, deformed}
monochromator/analyzer set that is used in almost all experiments.}
 & HFBS & SPHERES\\\hline
  reactor power (MW) & 58 & 20 & 20\\
  PG(002) speed (m/s) & 75 & 250 & 100\\
  Si(111) thickness ($\upmu$m) & 700 & 750 & 500\\\hline
  resolution fwhm ($\upmu$eV) & 0.8--0.9 & 0.93 & 0.62--0.65\\
  flux at sample\footnote
{While chopper open. Au foil activation gives half this value.}
 ($10^6$~s$^{-1}$) & 1.2 & 2.4 & 1.8\\
  rescaled flux\footnote{Flux at sample, divided by resolution width
and reactor power, arbitrarily normalized to the value of SPHERES} &
0.17 & 0.91 & 1 \\
  signal-to-noise ratio\footnote
{Best value reported for a non-absorbing sample of optimum thickness.}
 & 1000\footnote{Not mentioned in the instrument
descriptions\protect\cite{FrMB97b,FrGo01,Fri02,rsp:in16char},
but confirmed by instrument users.} & 600 & 1750 \\
  \hline\hline
  \end{tabular}
  \end{center}
\caption{\label{TComp}Comparison of the three backscattering spectrometers
IN16 of the ILL,\protect\cite{FrMB97b,FrGo01,Fri02,rsp:in16char}
HFBS at NIST,\protect\cite{MeDG03}
and SPHERES of the JCNS at FRM~II.}
\end{table}

The instrumental resolution is within 5\% proportional
to the chosen thickness of the Si(111) monochromator/analyzer wafers.
This confirms the fundamental importance of the crystal choice for
the instrument performance,
and it shows that all three instruments very closely attain
 the optimum resolution allowed by their crystal sets.

\textit{Ceteris paribus},
the flux at the sample should be proportional to the reactor power.
In a very good approximation,
it should also be proportional to the width of the energy
band selected by the monochromator.
Therefore, the figure of merit that describes the overall efficiency
of the cold source, of the neutron guides, and of the primary spectrometer
is flux divided by reactor power divided by resolution width.
The comparison of this
\textit{rescaled flux} shows how much the compact design of HFBS and SPHERES,
due to Schelten and Alefeld,\cite{ScAl84}
is superior to the double-deflector layout of IN16.

The primary spectrometer transmission is also somewhat related
to the signal-to-noise ratio:
to maximize the latter,
we installed in SPHERES \textit{ad hoc} several slits
that cut away some 10\% of the incoming and of the backscattered beam
in order to prevent neutrons from being transmitted through the
closed chopper.
The next chopper, currently under construction,
will hopefully allow us to take away these slits
and to increase the flux at the sample accordingly.
Furthermore, 
in this new chopper the PG(002) deflector crystals
will be moved with about optimum speed,
which will result in yet another important gain in flux.

At small scattering angles,
 analyzers cannot be aligned to exact backscattering geometry
because detectors must be placed outside the incoming neutron beam.
This makes the resolution 
much wider and more assymetric than in the regular backscattering detectors.
Supported by simulations,
we have improved the resolution at small angles
by breaking the azimuthal symmetry of the analyzer rings,
as will be described separately.

\section*{Acknowledgements}

Building and commissioning SPHERES has been funded by
the German Bundesministerium f\"ur Bildung und Forschung,
projects 05NX8CJ1 and 03RI16JU1.

Foundations for SPHERES were laid by former project scientists
Oliver Kirstein and Peter Rottl\"ander,
and by former project engineer Tadeusz Kozielewski.
Many other colleagues from Forschungszentrum J\"ulich
contributed to this project.
We would like to acknowledge
Ulrich Probst, Helga Straatmann and Thomas Koppitz
for development of the chopper rotor,
Ulrich Giesen and Ulrich Pabst for a vibrational analysis therof,
Christoph Tiemann for his contribution to the Doppler drive,
Gerd Schaffrath for the instrument housing,
Manfred Bednarek for the electrical installations,
Peter Stronciwilk and Marco G\"odel for mechanical constructions,
Vu Thanh Nguyen for the accurate glueing of the silicon crystals,
Harald Kusche, Andreas Nebel, and Bj\"orn Poschen for technical support,
and Alexander Ioffe for coordinating the JCNS outstation at FRM~II.
Sergej Manoshin and Alexander Ioffe
contributed a simulation of the convergent neutron guide.
Michaela Zamponi, the new instrument responsible since fall 2011,
contributed to the last Au activation measurement.

We are grateful to Winfried Petry, J\"urgen Neuhaus,
and all the staff of FRM~II for their hospitality.
We thank Harald T\"urck for generous help at all levels of engineering,
Christian Breunig for the neutron guide,
Philipp J\"uttner for the shutter construction,
Helmut Zeising, Birgit Wierczinski and their entire team
for smooth handling of the radioprotection necessities,
and Ralf Lorenz for sensibly supervising our safety.

We thank our colleagues and friendly concurrents
Bernhard Frick, Tilo Seydel, and Lambert van Eijck
of the ILL
for stimulating discussions and the open exchange of experiences and ideas.
Victoria Garc\'ia Sakai, Goran Gasparovic, and Dan A. Neumann
kindly showed us HFBS at NIST,
and Andreas Meyer provided insights from his commissioning experience.





\end{document}